# Making and identifying optical superposition of very high orbital angular momenta

**Lixiang Chen\*, Wuhong Zhang, Qinghong Lu, and Xinyang Lin**

*Department of Physics and Laboratory of Nanoscale Condensed Matter Physics, Xiamen University,*

*Xiamen 361005, China*

*\*Corresponding author: chenlx@xmu.edu.cn*

We report the experimental preparation of optical superpositions of high orbital angular momenta(OAM). Our method is based on the use of spatial light modulator to modify the standard Laguerre-Gaussian beams to bear excessive phase helices. We demonstrate the surprising performance of a traditional Mach-Zehnder interferometer with one inserted Dove prism to identify these superposed twisted lights, where the high OAM numbers as well as their possible superpositions can be inferred directly from the interfered bright multiring lattices. The possibility of present scheme working at photon-count level is also shown using an electron multiplier CCD camera. Our results hold promise in high-dimensional quantum information applications when high quanta are beneficial.

Recent years have witnessed a rapidly growing interest in twisted light in both classical and quantum regimes. Twisted light is so named because of its helical phase front of $\exp(i\ell\phi)$, where $\phi$ is the azimuthal angle and $\ell$ is an arbitrary integer [1]. The associated orbital angular momentum (OAM) eigenstates, $|\ell\rangle$, therefore form a complete and orthogonal basis, and the OAM quanta of $\ell\hbar$ per photon is theoretically discrete and unbounded [2]. Twisted light has been proved a very useful tool in many fields ranging from optical tweezers and spanner, microscopy and imaging, free-space communication, to quantum optics and quantum information [3, 4]. In all these applications, generation and measurement of twisted light play a fundamental role. Over the last two decades, various methods were established to produce twisted light, such as cylindrical lenses [5], spiral phase plates [6], spiral phase mirrors [7], inhomogeneous birefringent elements [8], and silicon-integrated optical vortex emitters [9]. Besides, the management of multidimensional OAM vector states is of significant importance in the generation of engineered qudits for high-dimensional quantum information applications [10, 11].

Here we make and identify superpositions of very high OAM carried by the modified Laguerre-Gaussian (LG) beams using computer-controlled spatial light modulator (SLM). Recently, SLM were widely employed to modulate the amplitude, phase, or polarization of a light beam [12]. Of particular interest is the elaborate design of specific digital holograms for preparing arbitrary OAM superpositions,which have been exploited for a deep insight into quantum correlations. The example applications include the demonstration of angular Einstein-Podolsky-Rosen correlation [13], violation of Bell-type inequalities [14], observation of entangled vortex links [15]. Besides, we have also used SLM to perform ghost angular diffraction[16], high-dimensional OAM entanglement concentration [17], and logical proof of Hardy's nonlocality [18]. In these demonstrations, photon pairs generated by spontaneous down-conversion carry a broad spectrum of OAM and these OAM values are low [19]. Based on entanglement transfer, photonic entanglement of very high OAM numbers was very recently reported [20], where the double superposition of opposite high OAM numbers were considered based on slit masks made by laser cutter that cuts the slits into the black paper. The main purpose of our work here focuses on the flexible preparation of high OAM superposition involving more OAM eigenmodes using SLM. We emphasize that our present work does not touch entanglement as in [20], while we use an electron multiplier CCD (EMCCD) camera to show the feasibility of our scheme working at photon-count level, such that the potential application in multidimensional entanglement is manifest. In our scheme, to make full use of the pixel arrays in SLM, here we modify the LG light beams to bear excessive phase helices. Based on a Mach-Zehnder interferometer with one inserted Dove prism, we further identify various superposition by the interfered bright multiring lattices.

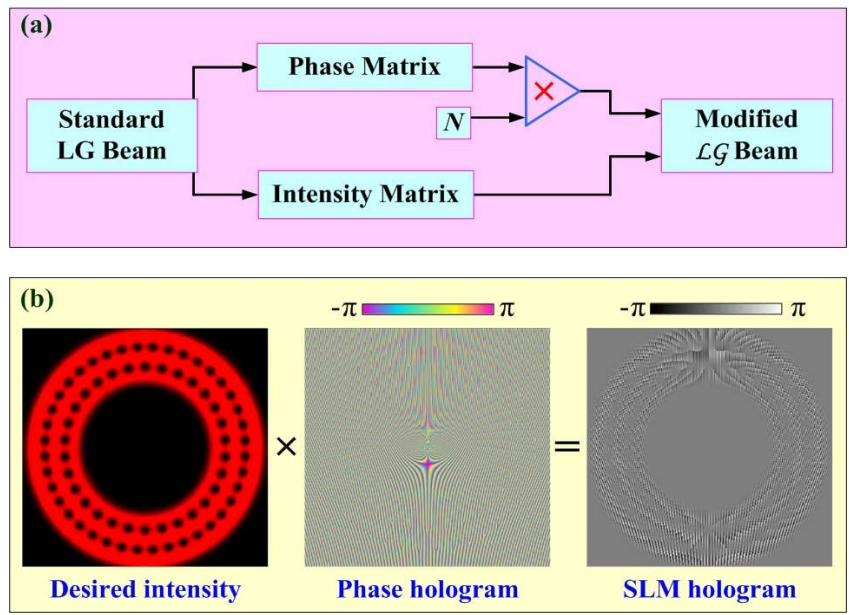

**FIG. 1.** (Color online) Illustration of algorithm: (a) for making the modified $\mathcal{LG}$ beam from a standard LG one; (b) for generating the SLM hologram.

The LG beams are considered as a natural choice for the description of twisted light carrying OAM [1]. In the cylindrical coordinate $(\rho, \phi)$, a standard LG mode at the beam waist plane $z = 0$ is described by,

$$LG_p^\ell(\rho,\phi) = R_p^\ell(\rho)\exp(i\ell\phi)\,,  \tag{1}$$

and,

$$R_p^\ell(\rho) = A_{p,\ell}(\frac{\sqrt{2}\rho}{\omega})^{|\ell|}\exp(\frac{-\rho^2}{\omega^2})L_p^{|\ell|}(\frac{2\rho^2}{\omega^2})\,,  \tag{2}$$

where $\omega$ is the beam waist, $A_{p,\ell}$ the normalized constant, $L_p^{|\ell|}(\cdot)$ the generalized Laguerre polynomial, and $R_{P,\ell}(\rho,\phi)$ describes the intensity distribution while $\exp(i\ell\phi)$ describes the helical phase structure, with $p$ and $\ell$ being the radial and azimuthal mode indexes, respectively. LG beams with the same $\ell$ but different $p$ indices carry the same OAM amount of $\ell\hbar$ per photon. Here we restrict out attention on those high-order LG modes with $p=0$, namely, of a single bright annular ring and zero on-axis intensity. Besides, in order to make full use of the pixel arrays of SLM, we make the modified $\mathcal{LG}$ beams rather than the standard ones to bear excessive phase helices, namely,

$$\mathcal{LG}_N^\ell(\rho,\phi) = R_{p=0}^\ell(\rho)\exp(iN\ell\phi)\,,  \tag{3}$$

Our algorithm is illustrated by Fig. 1(a). Such modified $\mathcal{LG}$ beams at the beam waist plane remains the same intensity distribution of single bright ring as the standard $LG_{p=0}^\ell$, but carry $N\ell\hbar$ OAM per photon, that is, $N$ times as high as the latter one. The propagation dynamics of the modified $\mathcal{LG}_N^\ell$ beams can be studied within the frame of standard LG beams, after making such a decomposition: $\mathcal{LG}_N^\ell(\rho,\phi) = \sum_p c_p LG_P^{N\ell}$, where $c_p$ denotes the overlap probability. However, we here focus only on their OAM content as well as their possible superposition with high OAM numbers. As is shown below, introduction of these $\mathcal{LG}_N^\ell$ beams facilitates the versatile use of our SLM to prepare the high OAM superposition that involves more OAM modes.

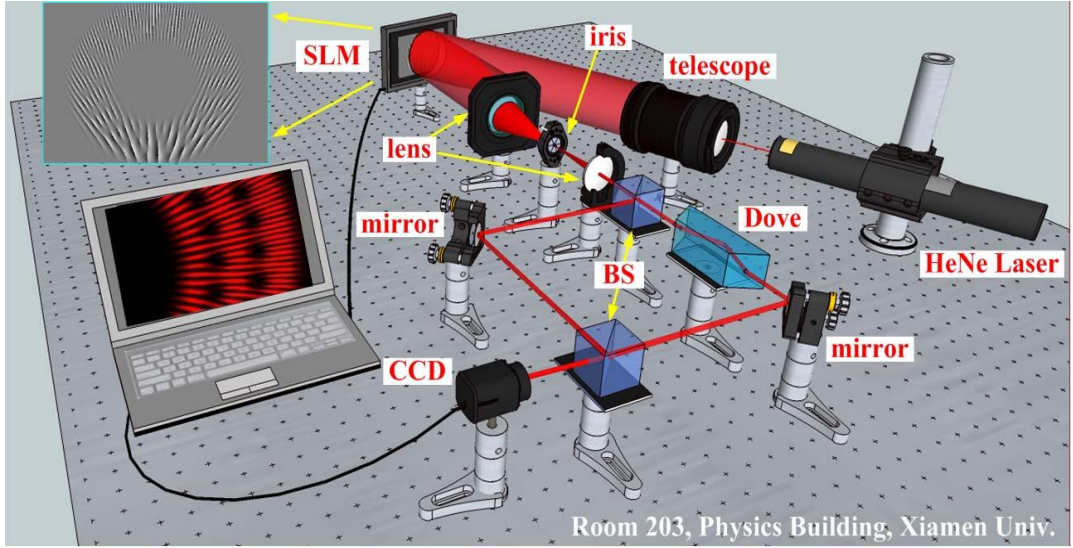

**FIG. 2.** (Color online) Optical system for making and identifying optical Superpositions of $\mathcal{LG}$ beams carrying high OAM.

Figure 2 shows a schematic overview of the experimental setup. The input light is a linearly polarized zero OAM Gaussian one derived from a 5 mW, 633 nm HeNe laser. After collimated by a telescope, the light beam is expanded and incident on a computer-controlled SLM (Hamamatsu, X10486-1). The SLM is a reflective device consisted of an array of pixels (792 × 600) with an effective area of 16 mm × 12 mm and a pixel pitch of 20 μm. Each pixel imprints individually the incoming light with a phase modulation (0 ~ 2π) according to the 8-bit grayscale (0 ~ 255). And the whole SLM acts as reconfigurable diffractive element, allowing an interactive manipulation with a response time comparable to the video displays. A frequently used design is to add a blazed grating modulo 2π to a spiral phase of $exp(i\ell\phi)$, then we obtain a forked hologram, whose first-order diffracted beam carries $\ell\hbar$ OAM per photon [21]. Although the SLM we use is phase only modulator, it can be utilized to shape the intensity of the diffracted beam also. As illustrated by Fig. 1(b), this is achieved by multiplying the phase hologram with the desired intensity distribution, and the final hologram addressed by SLM is given by [22],

$$\Phi(\rho,\phi)_{SLM} = [\Phi(\rho,\phi)_{Desired} + \Phi(\rho,\phi)_{Linear}]_{\mathrm{mod}2\pi} \times \mathrm{sinc}^2[1-\pi I(\rho,\phi)_{Desired}], \quad (4)$$

where $\Phi(\rho,\phi)_{Desired}$ and $I(\rho,\phi)_{Desired}$ are the desired phase and intensity distributions, respectively; $\Phi(\rho,\phi)_{Desired}$ is the phase of the linear grating, and $\mathrm{sinc}^2(\cdot)$ accounts for the mapping of the phase depth to the diffraction efficiency of the spatially dependent blazing function. Along this line, we aim to make virous superpositions of the modified $\mathcal{LG}$ modes, namely,

$$f(\rho,\phi) = \sum_{\ell} \alpha_{N,\ell} \mathcal{LG}_N^{\ell}(\rho,\phi), \quad (5)$$

where $|\alpha_{N,\ell}|^2$ characterizes the spiral spectrum. According to Eq. (4), we have $\Phi(\rho,\phi)_{Desired} = \arg(f(\rho,\phi))$ and $I(\rho,\phi)_{Desired} = |f(\rho,\phi)|^2$. In the quantum language, we can rewrite simply Eq. (5) as $|\varphi\rangle_N = \sum_{\ell} \alpha_{N,\ell} |\mathcal{LG}_N^{\ell}\rangle$ where the $\mathcal{LG}_N^{\ell}$ modes can be adopted as the OAM eigenstates. In order to produce high OAM effectively, we make the beam waist of the highest OAM component to match the outermost circumference of the SLM, such that more pixels per 2π phase shift are available. We estimate there are approximately 2666 pixels in the outermost circumference, which suggests for example only about 6 pixels per 2π phase change for $\ell = 400$. In [20], the double OAM superposition is measured using slit masks made by laser cutter that cuts the slits into the black paper. In contrast, our scheme makes good use of SLM to prepare reconfigurable diffractive holograms, therefore allowing an interactive and real-time manipulation.

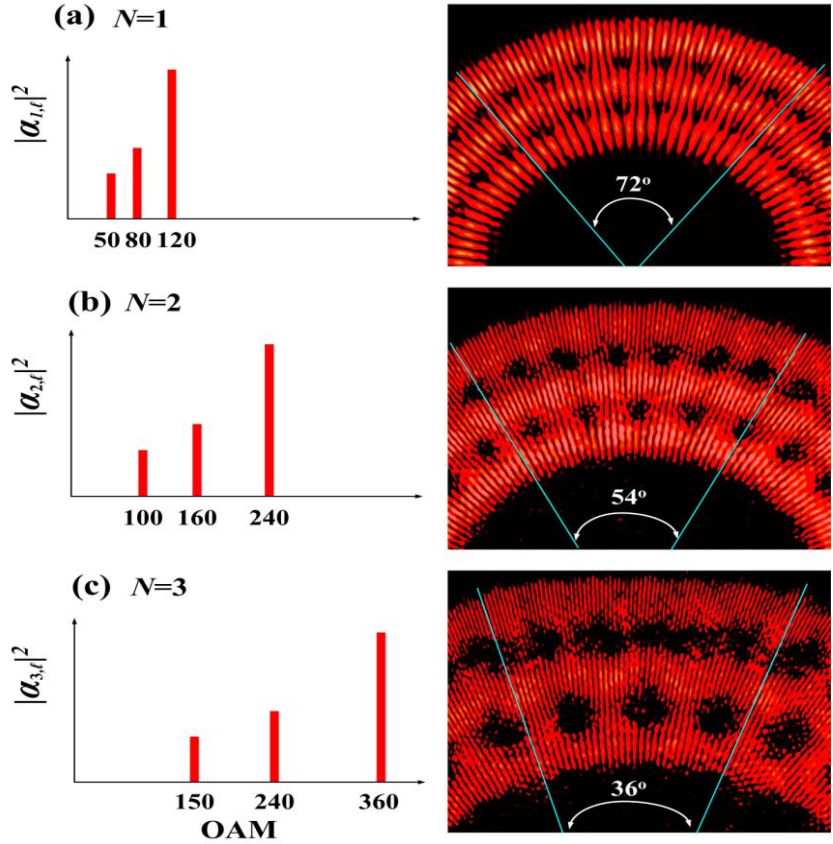

**FIG. 3.** (Color online) Left column shows the theoretical spiral spectra and right column presents the partial inferferograms of bright multiring lattices measured experimentally using color CCD camera: (a) $N = 1$, (b) $N = 2$, and (c) $N = 3$.

To verify that the coherent superposition described by Eq. (5) have been produced by SLM, we make the first diffraction order of reflected light propagate through a 4f system consisted of two lenses and an adjustable iris placed at the focal plane. Subsequently, the diffracted light is steered into a traditional Mach-Zehnder interferometer with one inserted Dove prism. We note that such interferometers are actually not new, for example, they or their modifications have been used extensively to reveal the phase structures of light or route OAM of photons [2, 23–27]. Here we focus further on its surprising capability to characterize very high OAM as well as their arbitrary superposition. By the first non-polarizing beam splitter (BS), the incoming light is divided equally and directed into two arms. The Dove prism inserted in one arm is used to flip the transverse cross section of the transmitted beam such that the OAM is reversed. The light components from two arms are recombined again in the second BS. A color CCD camera is used to monitors the output intensity, and the recorded interferogram can be described as a consequence of the interference between the incoming light and its mirror, namely,

$$I(\rho,\phi) \propto \left|\sum_{\ell} \alpha_{N,l} R_{p=0}^{\ell}(\rho)(\exp(iN\ell\phi)+\exp(-iN\ell\phi))\right|^2 = 2\left|\sum_{\ell} \alpha_{N,\ell} R_{p=0}^{\ell}(\rho)\right|^2 (1+\cos(2N\ell\phi)). \quad (6)$$

As can be seen from Eq. (6), complete constructive or destructive interference occurs at angle $\phi$ determined by $cos(2N\ell\phi)=1 \, or -1$, which indicates that interfering an $exp(iN\ell\phi)$ beam with its mirror image produces a pattern of $2N\ell$ radial spokes. Besides, as indicated by the term of $\left|\sum_{\ell} \alpha_{\ell} R_{p=0}^{\ell}(\rho)\right|^2$ in Eq. (6), we know the superposition of $\mathcal{LG}_N^{\ell}$ beams with different $N$ and $\ell$ indices results in an interferogram of multiple bright rings, as the radius of each ring scales with $\sqrt{\ell}$ [28]. And the ring brightness is determined by both the mode weight and the right radius, approximately given as $\left|\alpha_{N,\ell}\right|^2/\sqrt{\ell}$. Besides, the spoke number can be given by $n=2N\ell$. This suggests a way for us to acquire and identify inversely the information about the high OAM numbers and possible superposition by measuring and analyzing these interesting Interferograms.

By adopting the modified $\mathcal{LG}$ beams as OAM eigenstates, it is rather convenient to make very high OAM just by tuning N, without the need to change the beam waist each time to optimize the SLM hologram. We show our experimental results in Fig. 3, where the prepared superposition of high OAM can be described as in terms of $\mathcal{LG}$ modes, namely,

$$\left|\varphi\right\rangle_N = 0.68\left|\mathcal{LG}_N^{120}\right\rangle + 0.57\left|\mathcal{LG}_N^{80}\right\rangle + 0.46\left|\mathcal{LG}_N^{50}\right\rangle. \quad (7)$$

By changing N = 1, 2 and 3, we can then prepare easily the highest OAM numbers up to $\ell = 120$, 240 and 360, respectively. For each N, the hologram addressed by SLM has the similar profile to the optimal one shown in Fig. 1(b). They produce high OAM superposition of different phase structures but with the same intensity distribution, as mentioned above. The left column in Fig.3 shows the spiral spectra we prepared according to Eq. (7), and the right column shows correspondingly the interferograms we measured, which are in good agreement with the theoretical predictions from Eq. (6). For N = 1, see Fig. 3(a), the superposition is that of $\ell$ = 120, 80, 50, and there are three bright concentric rings, where the inner one consists of n = 100 spokes ($\ell = 50$), the middle 160 spokes ($\ell = 80$), and the outer 240 spokes ($\ell = 120$). The total spoke number of each ring can be calculated directly from the central angles $\theta$ (in degree) and the subtended spoke number $n_0$, namely, $n = 360 n_0/\theta$. As can be seen from Figs. 3(b) and 3(c), the spoke number of each ring for N = 2, 3 are two and three times of that for N = 1, respectively. These observations verify that the radius of each ring does scale with $\sqrt{\ell}$ while the spoke number $N = 2n\ell$. Therefore these clear interferograms confirm the effectiveness of our algorithm used for making $\mathcal{LG}$ beams, and the good quality of high OAM superposition we have made using SLM.

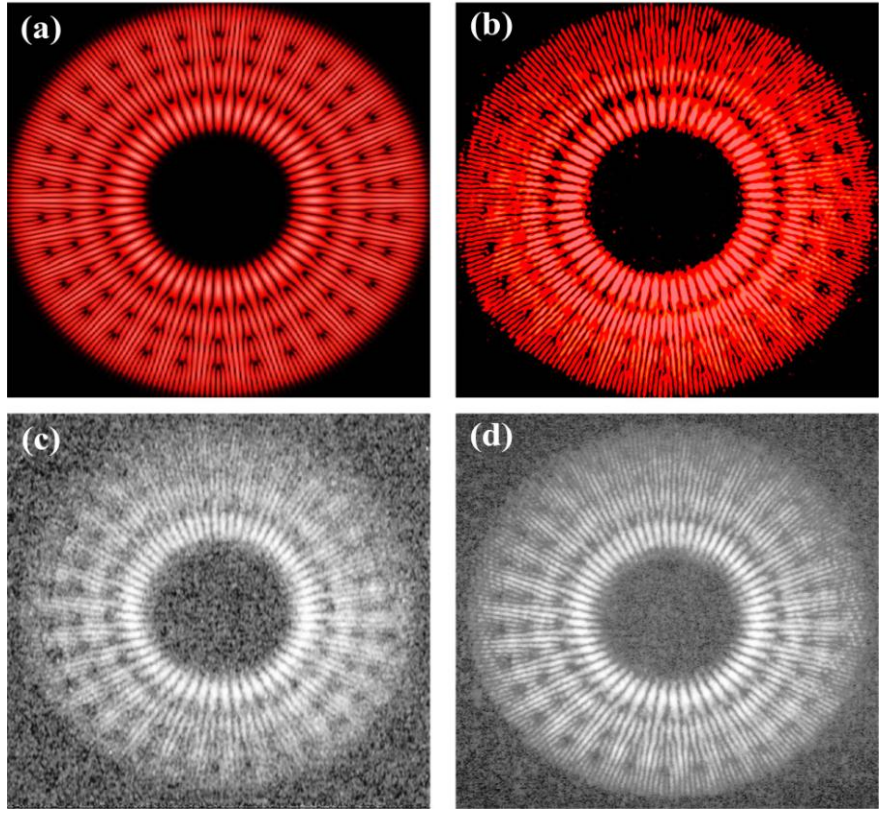

**FIG. 4.** (Color online) Quadruple superposition $|\varphi\rangle = (|120\rangle + |90\rangle + |60\rangle + |30\rangle)/2$. (a) is the numeric simulation, (b) is the experimental measurement using color CCD camera, (c) and (d) are those using EMCCD camera, where the average photon fluxes per pixel are $\approx 0.16$ and $1.00$, respectively.

Within the frame of quantum optics, the superposition of twisted light can be considered as the ideal candidate for manipulating multidimensional vector states residing in that high-dimensional OAM Hilbert space. To this end, we show the feasibility of our scheme to make a four-level OAM superposition (ququart), namely,

$$|\varphi\rangle = \frac{1}{2}(|120\rangle + |90\rangle + |60\rangle + |30\rangle), \qquad (8)$$

where $|\mathcal{LG}_{N=1}^{\ell}\rangle$ is labeled as $|\ell\rangle$ for short. The numerical simulation of the interfered pattern performed based on Eq. (6) is shown by Fig. 4(a), while the experimental result recorded by the color

CCD camera is presented by Fig. 4(b). There are four bright rings appearing concentrically, and the spoke number of each ring is just twice that of individual OAM. Another characteristic feature of the interferograms is the observation of chains of "fork" structures distributing over the transition zone between two neighboring bright rings in the interferograms. A careful study finds that they are just positioned at the points of phase singularity of the superposed light, and as a whole, forming the dark multiring lattices, see Fig. 4(a) and 4(b). In this sense, these generated bright multiring lattices can be used for trapping atoms in red-detuned light, and dark multiring lattices suitable for trapping atoms with minimal heating in the optical vortices of blue-detuned light [29]. To demonstrate further that our scheme can work also at the photon-count level, another experiment is carried out using the low-noise electron multiplier CCD (EMCCD) camera (E2V, L3C216, 768 x 288 pixels) instead of the color CCD camera. This is achieved by inserting a series of neutral-density filters to attenuate the laser power into a very faint intensity. In Fig. 4(c), the interfered spokes in the three inner rings can be distinguished well, but those in the outmost ring appear as a blur in the background noise. This is because in this case the average photon flux per pixel is as low as only $\approx 0.16$. For comparison, we increase the photon flux per pixel up to $\approx 1.00$ by removing some neutral-density filters, and we obtain an inferferogram of four concentric rings with clearer edge and better contrast, see Fig. 4(d).

The ability of our scheme working at photon-count level implies that it may be applied to explore the quantum correlations of entangled photon pairs, where they are useful for real-time imaging of entanglement [30, 31] or full field-of-view ghost imaging [32]. Besides, if the down-converted photons are directed backwards into our present interferometer and the HeNe laser is substituted by the single-photon detector, then our scheme enables the detection of much higher-dimensional OAM superposition states. For example, when the SLM hologram is prepared according to Eq. (7), the detected state $|\varphi\rangle_{6D} = 0.48|-120N\rangle + 0.40|-80N\rangle + 0.33|-50N\rangle + 0.40|80N\rangle + 0.48|120N\rangle$ and it is of six dimensions, where $|NL\rangle$ stands for $|\mathcal{LG}_N^\ell\rangle$ for short. While for Eq.(8), an eight-dimensional superposition state, $|\varphi\rangle_{8D} = \frac{1}{2\sqrt{2}}(|-120\rangle + |-90\rangle + |-60\rangle + |-30\rangle + |30\rangle + |60\rangle + |90\rangle + |120\rangle)$, could be detected. From the applied point of view, the advantages arising from these various superposition with more and high OAM quanta may be twofold. First, it increases the possibility towards the realization of the so-called macroscopic entanglement. Second, it improves the angular resolution in remote sensing. Also, our results hold promise for the implementation of multidimensional alphabets or engineered qudits in connection with quantum entanglement [10], such as quantum bit commitment [33] and quantum key distribution [34].

In conclusion, we have made a variety of high OAM superpositions based on the modified $\mathcal{LG}$ beams using SLM. The excellent performance of the Mach-Zehnder interferometer to identify these superposed twisted lights were demonstrated. We acquire the information of the high OAM numbers and their possible superposition directly from the characteristic interferograms of bright multiring lattices. The feasibility of our scheme working at photon-count level is shown using EMCCD camera, and the possibility exploited for high-dimensional quantum information application is manifest.

**Acknowledgments:** L.C. thanks Prof. Miles Padgett and his Optics Group at University of Glasgow for their kind help in the LabVIEW programming. This work is supported by the National Natural Science Foundation of China (NSFC) (11104233), the Fundamental Research Funds for the Central Universities (2011121043, 2012121015), the Natural Science Foundation of Fujian Province of China (2011J05010), the Program for New Century Excellent Talents in University of China (NCET), and the Program for New Century Excellent Talents in Fujian Province University.